\documentclass[aps, prd, twocolumn,superscriptaddress]{revtex4}
\usepackage{graphicx}
\usepackage{epstopdf}
\usepackage{dcolumn}
\usepackage{bm}
\usepackage[english]{babel}
\usepackage{enumitem}

\date{\today}

\begin{document}

\title{Vacuum particle-antiparticle creation in strong fields as a field induced phase transition}

\author{S.~A.~Smolyansky}
\author{A.~D.~Panferov}
\affiliation{Saratov State University, RU - 410026 Saratov, Russia}
\author{D.~B.~Blaschke}
\affiliation{Institute for Theoretical Physics, University of Wroclaw, 50-204 Wroclaw, Poland}
\affiliation{Bogoliubov Laboratory for Theoretical Physics, Joint Institute for Nuclear Research, RU - 141980 Dubna, Russia}
\affiliation{National Research Nuclear University (MEPhI), RU - 115409 Moscow, Russia}
\author{L.~Juchnowski}
\affiliation{Institute for Theoretical Physics, University of Wroclaw, 50-204 Wroclaw, Poland}
\author{B.~K\"ampfer}
\author{A.~Otto}
\affiliation{Helmholtz-Zentrum Dresden-Rossendorf, 
D-01314 Dresden, Germany}
\affiliation{Institut f\"ur Theoretische Physik, TU Dresden, D-01062 Dresden, Germany}

\date{\today}

\begin{abstract}
The features of vacuum particle creation in an external classical field are studied for simplest external field models in $3 + 1$ dimensional QED.
The investigation is based on a kinetic equation that is a nonperturbative consequence of the fundamental equations of motion of QED.
The observed features of the evolution of the system apply on the qualitative level also for systems of other nature and therefore are rather general.
Examples from cosmology and condensed matter physics illustrate this statement.
The common basis for the description of these systems are kinetic equations for vacuum particle creation belonging to the class of integro-differential equations of  non-Markovian type with fastly oscillating kernel.
This allows to characterize processes of this type as belonging to the class of field induced phase transitions.
\end{abstract}


\maketitle

\section{Introduction \label{sect:in}}

In the  present work we investigate the features of the 
transition from an initial state of primordial vacuum oscillations to a final quantum field system of particles and antiparticles due to an external field (the dynamical Schwinger effect)
as a field induced phase transitions (FIPT).
As an example we consider a $3+1$ dimensional QED system in the presence of a linearly polarized time dependent electric field.
This particular case allows a rather simple kinetic description in the framework of the quasiparticle representation.
The corresponding kinetic equation (KE) has a specific structure: 
it is an integro-differential equation of non-Markovian type with a fastly oscillating kernel describing the evolution of vacuum oscillations excited by the external field.
The mathematical structure of this KE is preserved also for other systems with unstable vacuum.
Therefore one can expect that the investigated features of FIPT can be recognized also in other 
quantum field systems of corresponding nature in the presence of a strong classical external field.

The basic KE is given in Sect.~\ref{sect:1}.
Its numerical solutions are discussed in Sect.~\ref{sect:2} for some simple models for the external electric field.
The characteristic features in the description of the evolution of the particle-antiparticle plasma  created from vacuum are summarized in Sect.~\ref{sect:3}.

\section{Kinetic equation \label{sect:1}} 

According to the general theory of systems with unstable vacuum (see, e.g., \cite{Fradkin_1991, Gavrilov_1996}), two formulations of kinetic theory exist which are destined for the space-time description of vacuum exitations.
These are the KE in the Wigner \cite{Bialynicki-Birula_1991, Elze_1989, Hebenstreit_2010} and in the quasiparticle  \cite{Grib_1994, Schmidt_1998, Fedotov_2011} representations.
In Refs.~\cite{Fedotov_2011, Hebenstreit_Diss} the equivalence of these two approaches was demonstrated for the simplest external field models in QED.
For these simple, spatially homogeneous field models both versions of the kinetic theory allow for a reduction of the complicated  original equations to a rather simple system of ordinary differential equations (ODE) for three functions: the distribution function $f(\mathbf{p},t)$ of the quasiparticle excitations and two auxiliary functions for the description of the vacuum polarization.
The corresponding KE \cite{Schmidt_1998} describing vacuum electron-positron plasma (EPP) creation in a homogeneous linearly polarized electric field $E(t)=-\dot{A}(t)$ with the vector potential (in the Hamilton gauge) $A^\mu(t)=(0,0,0,A(t))$ is
\begin{equation}\label{ke}
\dot f(\mathbf{p} ,t) = \frac{1}{2} \lambda(\mathbf{p}
,t )\int\limits^t_{t_0} dt^{\prime} \lambda(\mathbf{p} ,t^{\prime})[1-2f(\mathbf{p}
,t^{\prime})]\cos\theta(t,t^{\prime}),
\end{equation}
where
\begin{eqnarray}
 \lambda(\mathbf{p},t) &=& e E(t)\varepsilon_{\bot}/\varepsilon^{2}(\mathbf{p},t), \label{lambda}
\\
\theta(t,t^{'}) &=& 2 \int^t_{t^{`
}} d\tau \, \varepsilon (\mathbf{p} ,\tau). \label{phase}
 \end{eqnarray}
Here $\lambda$ is the amplitude of the vacuum transitions, and $\theta$ is the high-frequency phase, describing the vacuum oscillations which are modulated by the external field. Furthermore, the quasienergy $\varepsilon$, the transverse energy $\varepsilon_\bot$ and the longitudinal quasi-momentum $P$ are defined as
\begin{eqnarray}
\varepsilon(\mathbf{p} ,t) &=& \sqrt{\varepsilon^2_{\bot}(\mathbf{p}) + P^2} ,
\label{eq:energy}
\\
\varepsilon_\bot &=& \sqrt{m^2 + p^2_\bot},
\label{eq:energy_perp}
\\
 P &=& p_\parallel -eA(t).
\label{eq:p-long}
\end{eqnarray}
Here $p_\bot=|\mathbf{p_\bot}|$ is the modulus of the vector $\mathbf{p_\bot}$ perpendicular to the field vector and  $p_\parallel=p_3$ is the momentum component parallel to the field.

The quasiparticle distribution function $f(\mathbf{p} ,t)$ is zero in the in-vacuum state where the external field strength is zero ($E_{\rm in}=0 $ corresponds to $A_{\rm in}=A(t_0)$), i.e. Eq.~(\ref{ke}) is complemented by the initial condition $f(\mathbf{p},t_0)=f_{\rm in}=0$.

It is assumed also that the electric field is switched off in the out-state ($E_{\rm out}=E(t\to\infty)=0$ and $A_{\rm out}\neq A_{\rm in}$). Thus, the in- and out-vacuum states are different.

The non-Markovian integro-differential equation (\ref{ke}) is equivalent to a system of three time-local ordinary differential equations
\begin{eqnarray}\label{ode}
 \dot{f} = \frac{1}{2}\lambda u, \quad \dot{u} = \lambda (1-2f) - 2 \varepsilon v, \quad \dot{v}= 2 \varepsilon u ,
\end{eqnarray}
where $u(\mathbf{p},t), v(\mathbf{p},t)$ are auxiliary functions describing vacuum polarization effects. 
The dynamical system (\ref{ode}) has the following integral of motion:
\begin{equation}\label{int}
    (1-2f)^2 +u^2+v^2=1
\end{equation}
compatible with the initial conditions $f_{\rm in}=u_{\rm in}=v_{\rm in}=0$.

In the low density approximation $2f \ll 1$, the KE (\ref{ke}) has a closed formal solution in the form 
of a useful quadrature formula  \cite{Schmidt_1999}
\begin{equation}\label{ld}
f(\mathbf{p} ,t) = \frac{1}{2}\int\limits^t_{t_0} dt^{\prime} \lambda(\mathbf{p}
,t^{\prime} )\int\limits^{t^{\prime}}_{t_0} dt^{\prime \prime} \lambda(\mathbf{p} ,t^{\prime \prime})\cos\theta(t^{\prime},t^{\prime \prime})~.
\end{equation}

The total number density of pairs is defined as
\begin{equation}\label{dens}
    n(t) = 2 \int \frac{d\mathbf{p}}{(2 \pi)^3} f(\mathbf{p} ,t)~,
\end{equation}
where the factor 2 corresponds to the spin degree of freedom.

In the present work the KE  (\ref{ke}) is solved numerically for two relevant models of the electric field: 
\begin{itemize}
\item[(i)] the Eckart-Sauter field with characteristic duration of action $T$
\begin{equation} \label{field2}
E(t) = E_0 \cosh^{-2}(t/T),  \,  A(t)= -{TE_0} \tanh(t/T) ,
\end{equation}
and 
\item[(ii)] the Gaussian envelope model of the laser pulse \cite{Alkofer_2009}
\begin{eqnarray} \label{field3} 
E(t) & = & E_0  \cos{ (\omega t) }\ e^{-t^2/2\tau^2 }, \\
A(t) &=& -\sqrt{\frac{\pi}{8}} E_0\tau \exp{(-\sigma^2/2)}\;\text{erf}\left(\frac{t}{\sqrt{2}\tau} -i\frac{\sigma}{\sqrt{2}}\right) + c.c. , \nonumber
\end{eqnarray}
where $\sigma= \omega \tau$ is a dimensionless measure for the characteristic duration of the pulse $\tau$ 
connected with the number of periods of the carrier field. 
\end{itemize}

The Eckart-Sauter field (\ref{field2}) admits an exact solution of the problem \cite{Grib_1994, Fedotov_2011, Nikishov_1970}; 
it is a benchmark case. 

In order to introduce the Keldysh parameter $\gamma = E_c \omega/E_0 m$ for the discussion of the field model (\ref{field2}) one can use the substitution $\omega \to 1/T$ .
In the limiting case $\gamma \ll 1$ the tunneling mechanism 
(with participation of an infinite number of photons) 
dominates, whereas for $\gamma \gg 1$ pair creation is driven by the absorption of few photons.

The vacuum oscillations (Zitterbewegung) play a crucial role in the mechanism of  vacuum EPP creation. 
The usual energy of vacuum oscillations $\varepsilon_0 = \sqrt{m^2 + \mathbf{p}^2}$ is transformed here to the quasienergy (\ref{eq:energy})
in the presence  of the time dependent electric field. 
The memory effect (non-Markovian character of the KE), 
the fastly oscillating factor with the phase (\ref{phase}) and the 
frequency $2\varepsilon$ (the dynamical energy gap) are the essential elements in the KE (\ref{ke}).
This equation contains two characteristic time scales: 
a slow one associated with the time scale of the external field period, $2\pi/\omega$, 
and a fast one given by the Compton time $\tau_c = 2\pi/m$. 
These scales are usually vastly different, $\omega \ll m$. 
The coupling of the dynamics related to these two scales leads to a very complicated structure of the distribution function, both in the first stage (generation of the quasiparticle EPP (QEPP)) and in the final stage (formation of the residual EPP (REPP)) \cite{Blaschke_2013}.

\section{Field induced phase transition \label{sect:2}}
In the considered situation, the FIPT appears as rearrangement of the vacuum state under the action of a classical electromagnetic field.
It leads to the $t -$ noninvariant quasiparticle vacuum which corresponds to a non-stationary Hamiltonian of the system (the S.Coleman theorem \cite{ Grib_1994, Coleman_1966}).
In this connection, the quasiparticle electron-positron pairs are the massive analog of the Goldstone bosons \cite{ Grib_1994, Glashow_1968}.

Let us consider phenomena which accompany the FIPT.

\subsection{Transient stage}

\begin{figure*}
\includegraphics[width=0.48\textwidth]{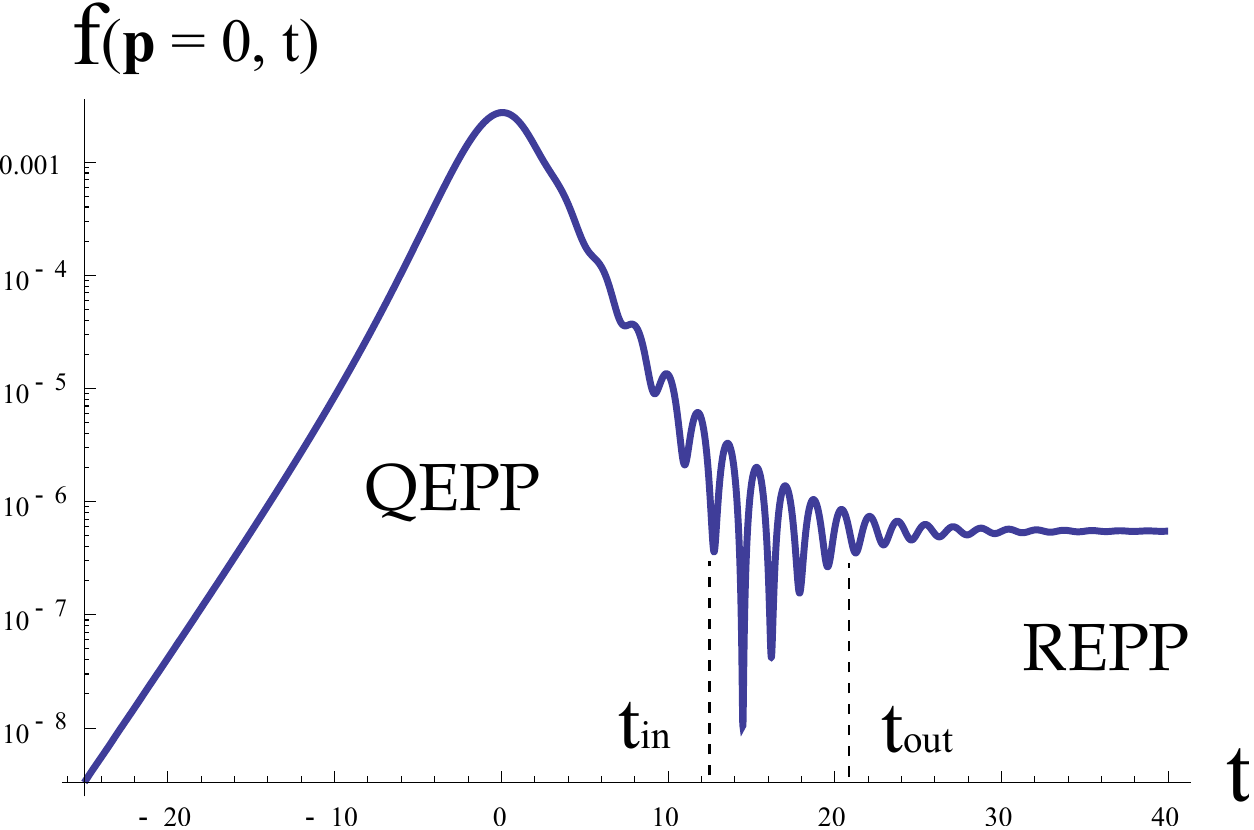} \hfill
\includegraphics[width=0.48\textwidth]{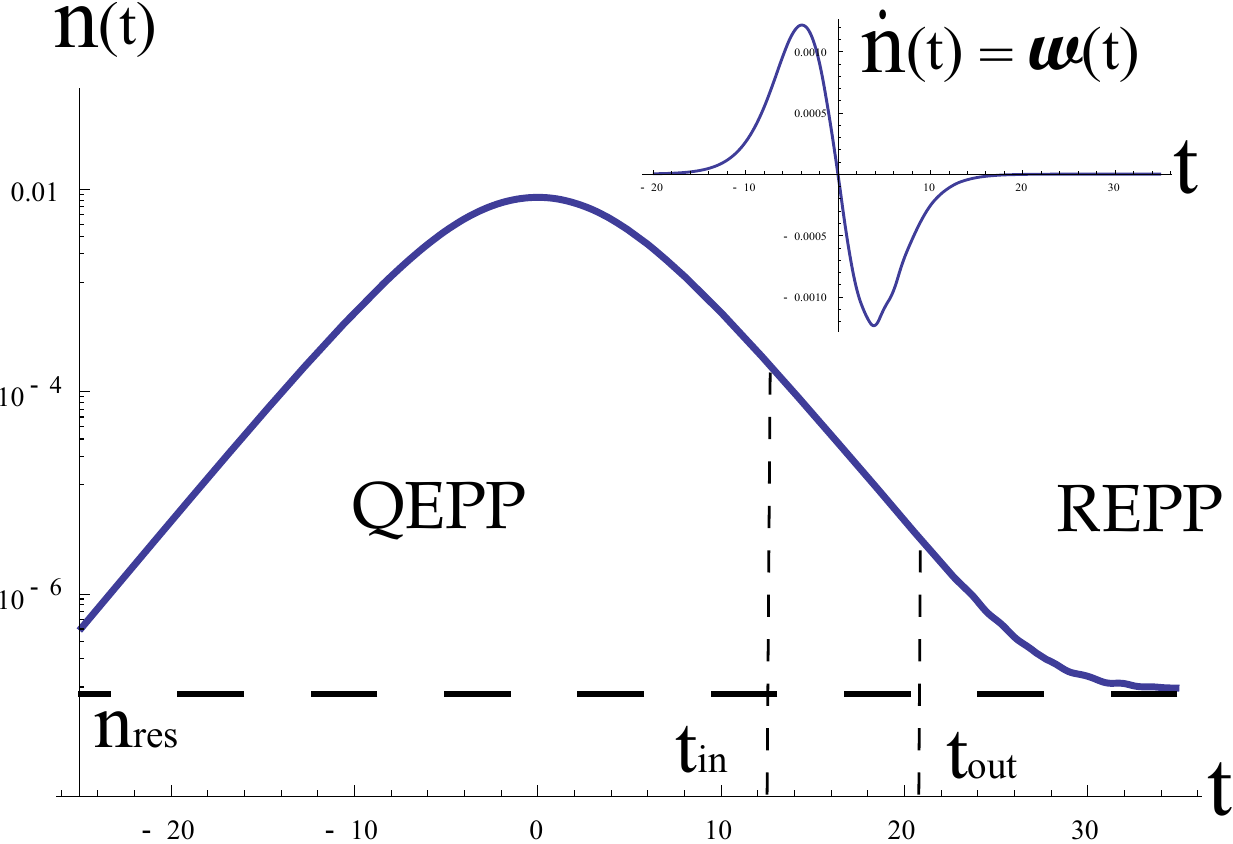}
\caption{The transition from the QEPP plasma to the final state for the Eckart-Sauter pulse type (\ref{field2}) 
with $E_0 = 0.2~E_c$ and $T = 8$. 
The labels $t_{in}$ and $t_{out}$ denote approximately the begin and the end of the transient stage.
{\bf Left panel:} Evolution of the distribution function for the point $p_\bot =  p_\parallel = 0$.
{\bf Right panel:} Evolution of the pair number density (\ref{dens}) and the local pair production rate 
$w(t) = \dot n(t)$ (inset). 
\label{fig:1}}
\end{figure*}

The typical picture of the EPP evolution under the action of the smooth pulse (\ref{field2}) is presented in Fig.~\ref{fig:1}. 
The left panel shows that the transient process of the fast EPP oscillations divides the evolution of the EPP into two domains, the QEPP and the REPP. 
After momentum integration the fast oscillations of the transient process are smoothed out, see the right panel of Fig.~\ref{fig:1}. 
The inset of that panel shows the local production rate. 
The results of the numerical solutions of the KE (\ref{ke}) (or (\ref{ode})) coincide with the exact solution \cite{Grib_1994, Nikishov_1970, Hebenstreit_Diss}. 
On all figures the time and frequency are scaled with the electron mass.

For qualitative orientation one can introduce here the time interval of the strong oscillations limited by point $t_{in}$ of the begin (that can be identified with the moment when the oscillations of the distribution function reach for the first time the level of the REPP) and the end $t_{out}$ (corresponding to the moment when the mean level of oscillations approaches that of the REPP and the elongation of the oscillations is significantly reduced). 
This transient period of the Zitterbewegung separates the smoothed QEPP stage from the REPP stage.

\begin{figure*}
\includegraphics[width=0.48\textwidth]{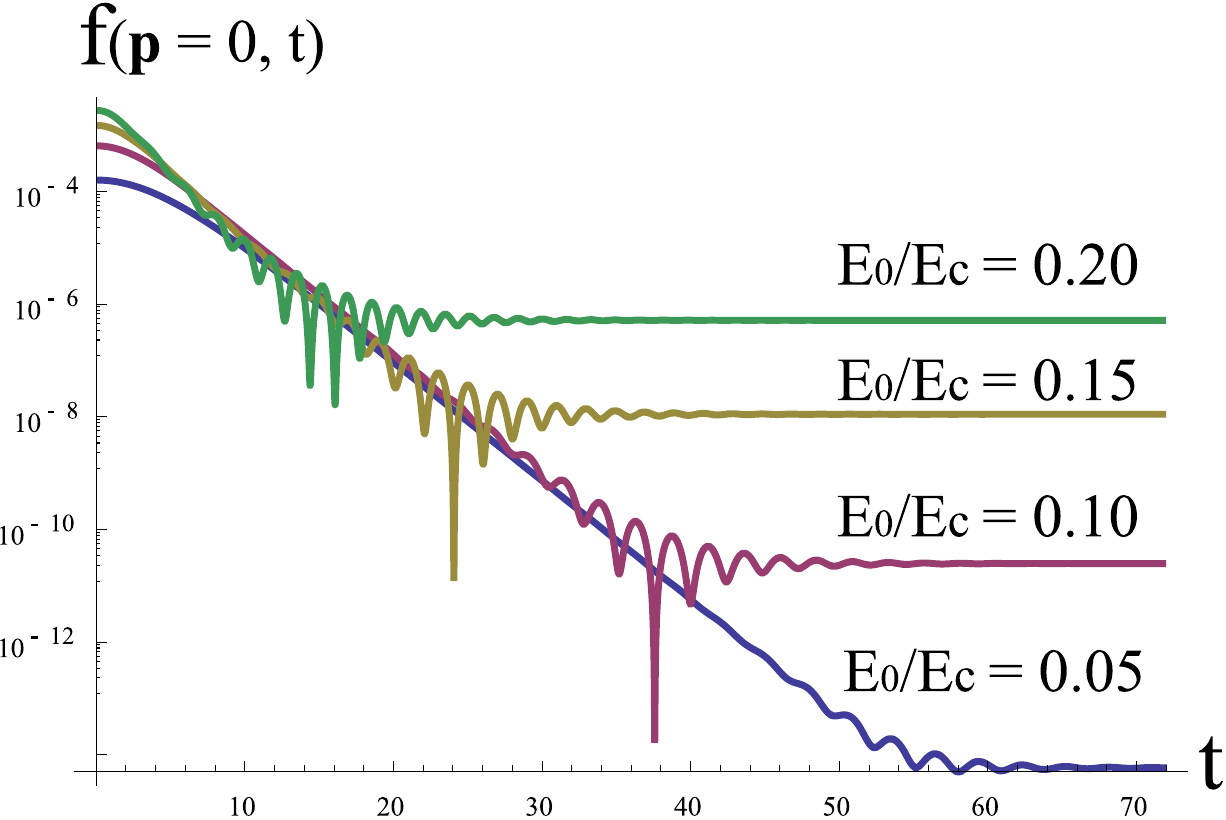} \hfill
\includegraphics[width=0.48\textwidth]{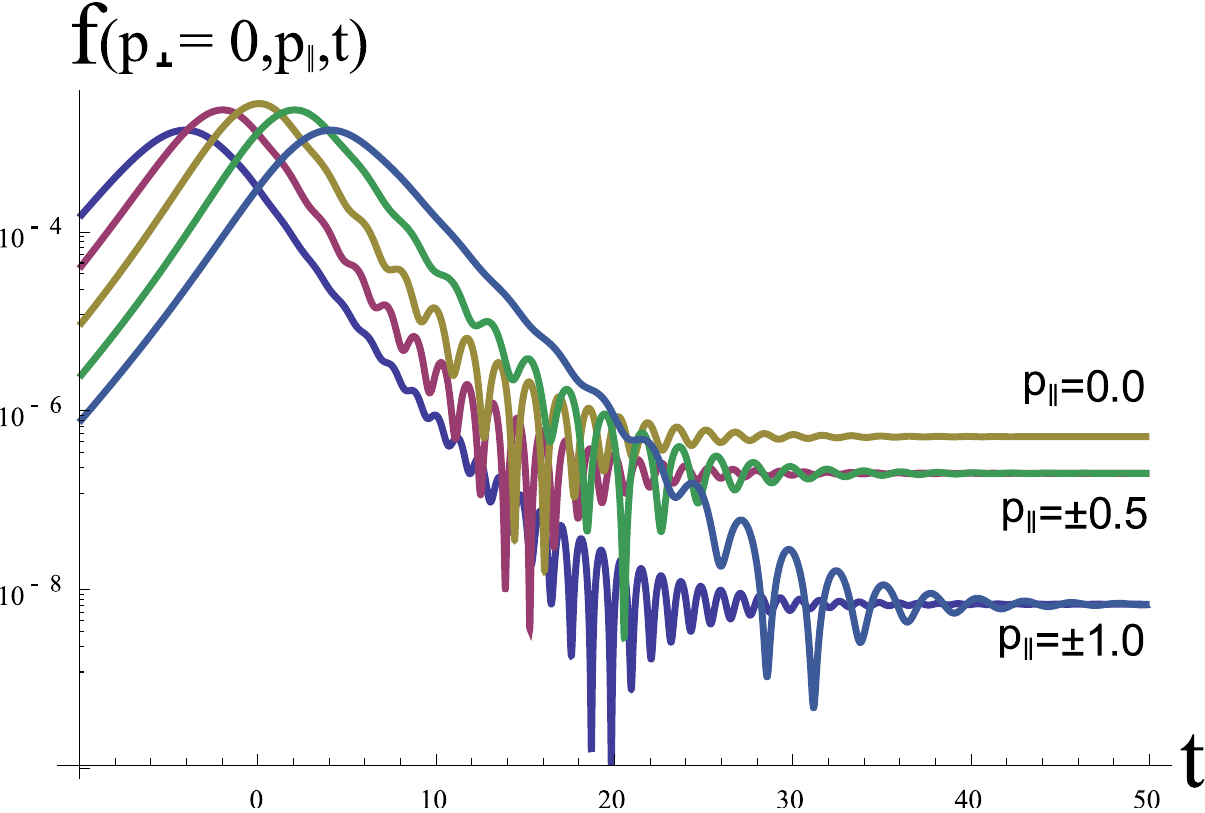}
\caption{Time evolution of the distribution function for the Eckart-Sauter pulse (\ref{field2}) with $T = 8.24$. 
{\bf Left panel:} At $p_\bot =  p_\parallel = 0$ for subcritical fields $E_0/E_c = 0.05, 0.10, 0.15$  and $0.20$. 
{\bf Right panel:} The details of phase transition at $E_0/E_c=0.2$ and $p_\bot =  0$ for different values  $p_\parallel = -1.0, -0.5, 0.0, 0.5, 1.0$.
\label{fig:2_}}
\end{figure*}

Under similar conditions strong oscillations are observed also in other physical models with massive constituents. 
For example, they appeared in the domain of the relativistic phase transition with dynamical mass generation (the inertial mechanism of particle creation) including the Higgs mechanism \cite{Toneev_2008}. 
Their existence can be found also in the strong field dynamical models of strongly correlated systems 
(see, e.g., Ref.~ \cite{SSP_3}).
Let us underline that the appearance of the transient region with strong oscillations takes place in the considered case of a smooth impulse  (\ref{field2}) without a carrier wave that would possess a high frequency component.

\begin{figure*}

\includegraphics[width=0.48\textwidth]{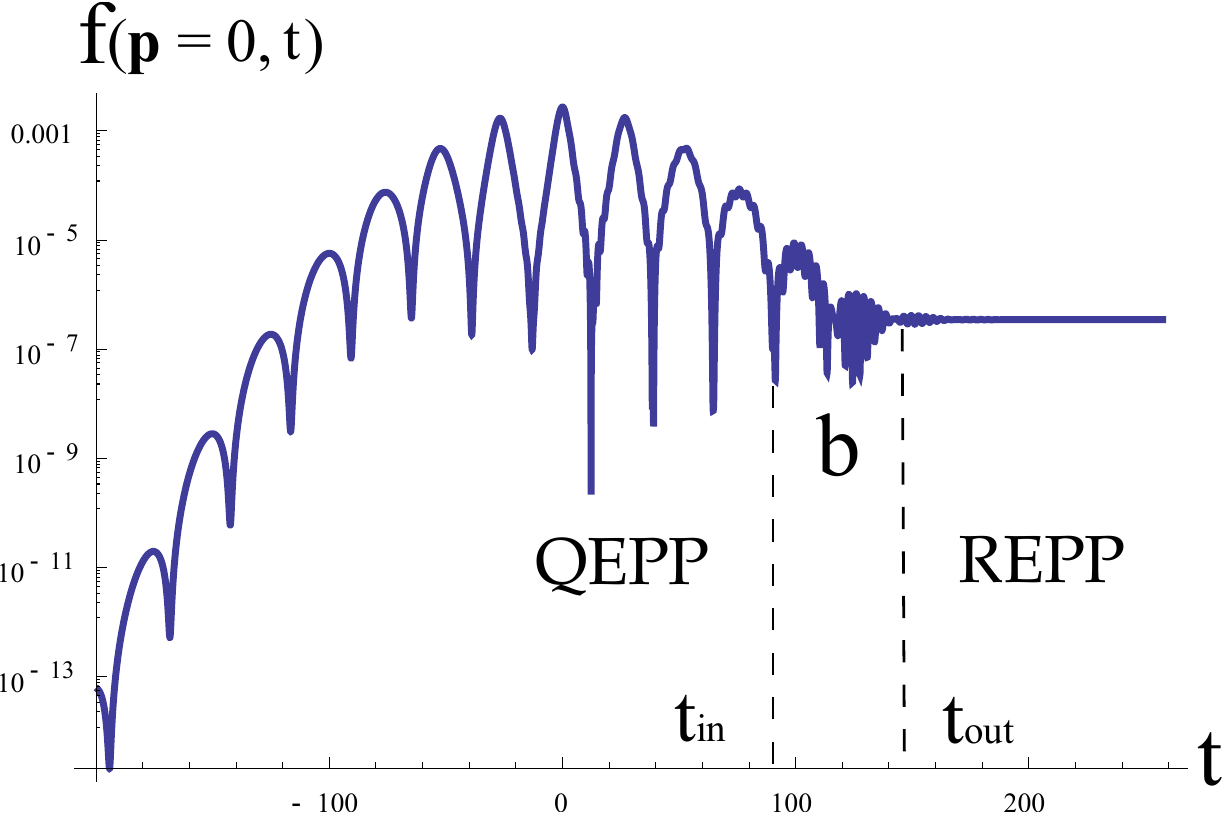} \hfill
\includegraphics[width=0.48\textwidth]{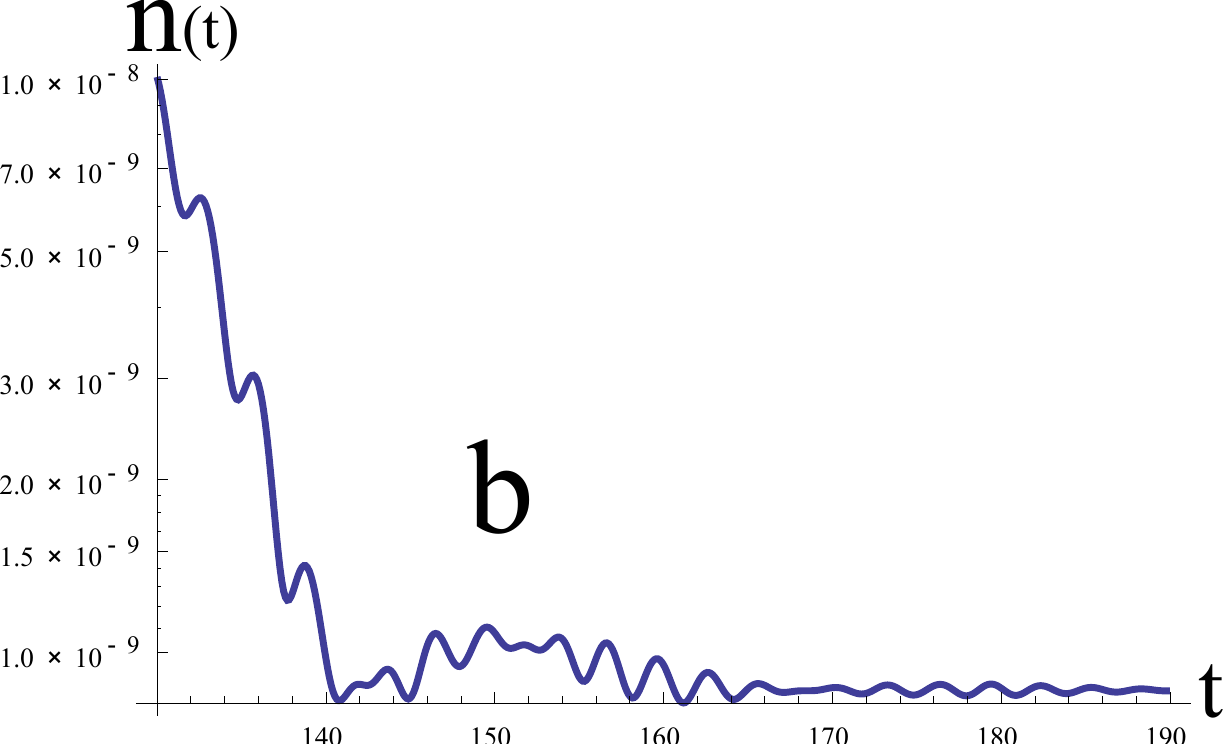}

\caption{Transition from QEPP to REPP in the case of a harmonic field with Gaussian envelope (\ref{field3}) with $\sigma = 5.0$.
{\bf Left panel:} The distribution function for the point $p_\bot =  p_\parallel = 0$. {\bf Right panel:} The density $n(t)$ (\ref{dens}) in the transition region $b$. 
\label{fig:3_}}
\end{figure*}

For a better understanding of this phenomenon let us consider the mechanisms of particle creation acting in the KE (\ref{ke}) or in its approximate solution  (\ref{ld}). 
We will trace the evolution of the system in the smooth field (\ref{field2}) for $t>0$ which is accompanied by a field strength depletion. 
If the electric field is rather strong, for $t<t_{\rm in}$ the acceleration mechanism represented by the force factor $eE(t)$ in the numerator of the amplitude (\ref{lambda}) is dominant whereas the fastly oscillating factor $\cos\theta(t,t^{\prime})$ on the r.h.s. of the KE (\ref{ke}) smoothes out. 
The vicinity of the moment $t_{in}$ of the begin of the transient stage is characterized by a weakening of the accelerating field action and by the growth of the role of the fast oscillations with the frequency $2\varepsilon(\mathbf{p} ,t) \ge 2m$, in which one can neglect now the influence of a weak field so that the oscillation ``beard'' in the transient stage appears  Fig.~\ref{fig:1}.
The subsequent field depletion accompanied by the growth of the vector potential (and the quasi-momentum $P(t)$ (\ref{eq:p-long})) in the denominator of the amplitude (\ref{lambda}) leads to the asymptotic extinction of the oscillations and the approach of the final REPP state.

Fig.~\ref{fig:2_} demonstrates the fine structure of the distribution function in the transient period for varying field strength at fixed momentum (left panel) and for varying $p_\parallel = p_3$ at $p_\bot = 0$ and fixed field strength (right panel).
One can see from here that the behavior of the distribution function depends on the selection of a point 
$\mathbf{p}$ in momentum space. 
The maximum of the distribution function is realised for $p_\bot = 0$ in different points $p_\parallel = p_3$ at different time moments because the component $p_\parallel$ is contained in the amplitude (\ref{lambda}) together with the vector potential in the quasi-momentum $P(t)$ (\ref{eq:p-long}). 
The features of these oscillations are defined by the double quasienergy $2\varepsilon(\mathbf{p} ,t)$ and are reproduced well also by the numerical solution of  the KE. 

These features of the transient process are complicated in the case of a high frequency periodic field with a Gaussian envelope (\ref{field3}), $\omega \gg 1/\tau$. 
Typical patterns are presented in Fig.~\ref{fig:3_}. 
In this case the transient stage separates into domains defined by the subcycle structure of the pulse. 
They are traced well also on the density curve of the EPP in the right panel of Fig.~\ref{fig:3_}, showing the last domain of the stage "b". 
Apparently, the effect of mutual amplification of EPP production as a result of the nonlinear interaction of the fast and slow components \cite{Dunne_2009,  Otto_2015, Otto_2016} of the field (\ref{field3}) is illustrated here.

A new element now is the dependence on the carrier frequency $\omega$. 
When comparing with the right panel of Fig.~{\ref{fig:1}}  we observe now the modulation effect which becomes apparent also in the area of the ``beard". 
The previously discussed picture of the transient regime occurs also at the end of each cycle of sub-pulses with a half-period duration $\pi/\omega$ of the external field  (\ref{field3}), but it gets squeezed by the neighboring cycle.
The general transient process arises at sufficient depletion of the envelope amplitude at $t > 0$.

\begin{figure*}
\includegraphics[width=0.48\textwidth]{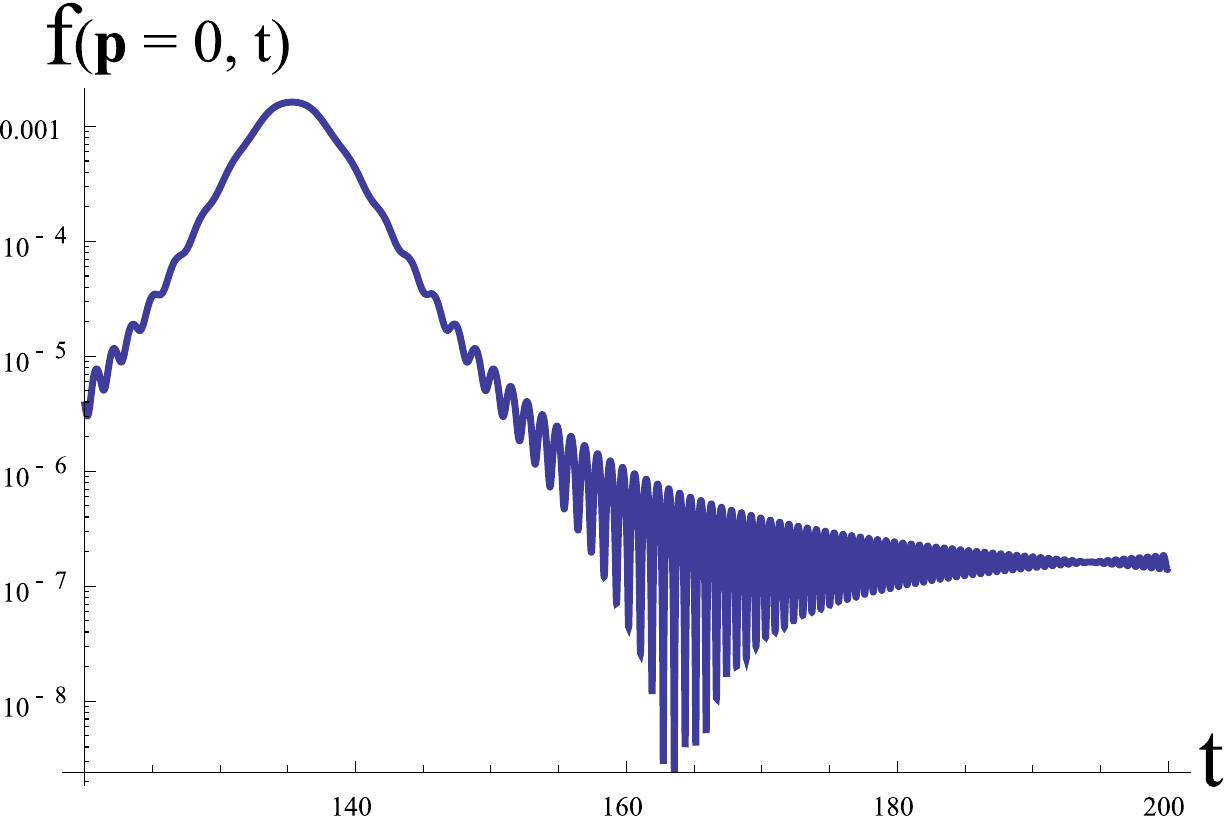} \hfill
\includegraphics[width=0.48\textwidth]{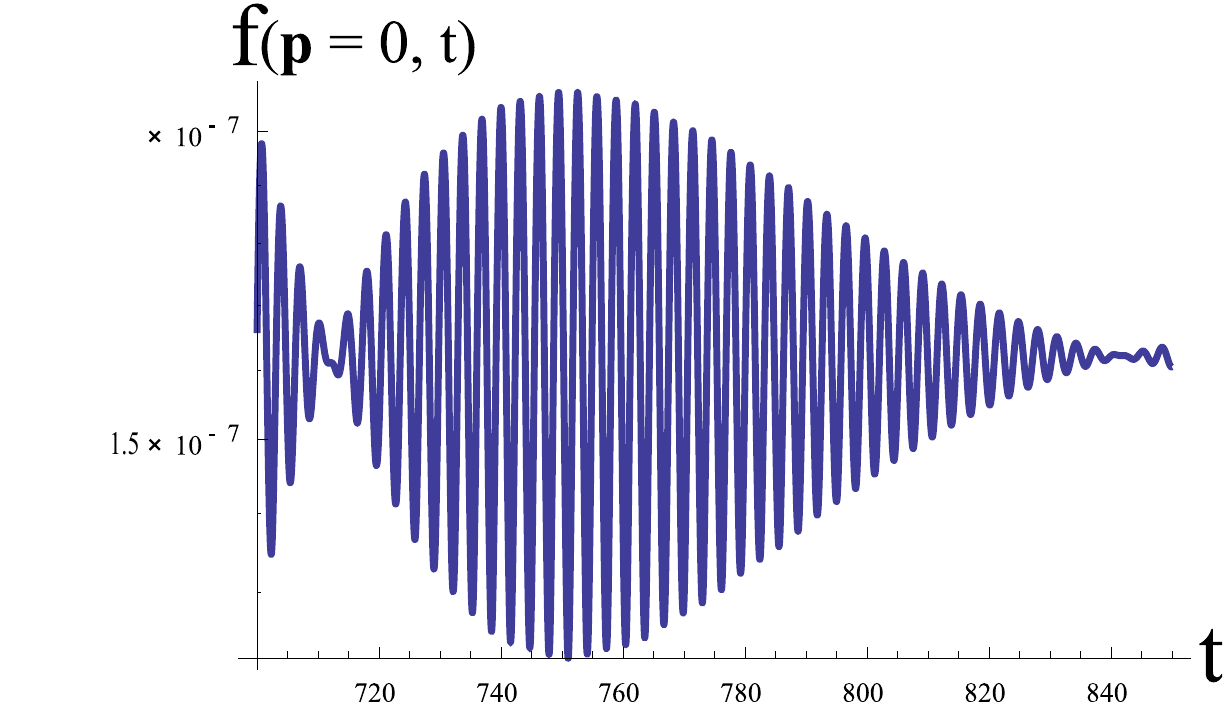}
\caption{Details of the evolution of the distribution function in the case of a harmonic field with the Gaussian envelope (\ref{field3}) with $\sigma = 5.0$ and wavelength $0.1$ nm.
{\bf Left panel:} Evolution of the EPP under the action of a single subpulse with a half-period duration. 
{\bf Right panel:}  The fine structure of the transient bubble (similar to Fig.~\ref{fig:3_}, left). 
\label{fig:5}}
\end{figure*}

On the left panel of Fig.~\ref{fig:5} we demonstrate this aspect of the evolution of the QEPP under the action of a single sub-pulse. 
For clarity of the picture we have increased here the wavelength in comparison with the case shown in 
Fig.~\ref{fig:3_}. 
Each such sub-pulse in the EPP evolution comprises all the quasiparticle stages of evolution, its transient stage and results in a partial REPP. 
The next sub-pulse will start with some nonzero EPP distribution.
This leads to the accumulation of REPP in the final out-state. 
The structure of the final transient region (the fragment ``b" on the Fig.~\ref{fig:3_}, left) is shown on the right panel of Fig.~\ref{fig:5}.

The presence of a transient region of fast oscillations in the distribution function is characteristic for every field model. 
In this regard the discussed phase transition under the action of a strong electric field is a universal effect for quantum field systems with an energy gap.
We remark that in the case of massless 2+1 dimensional QED (e.g., for graphene), the high-frequency transient region is absent and the evolution of the particle-antiparticle plasma distribution function is smooth \cite{Panferov_2017}.

\begin{figure*}
\includegraphics[width=0.44\textwidth]{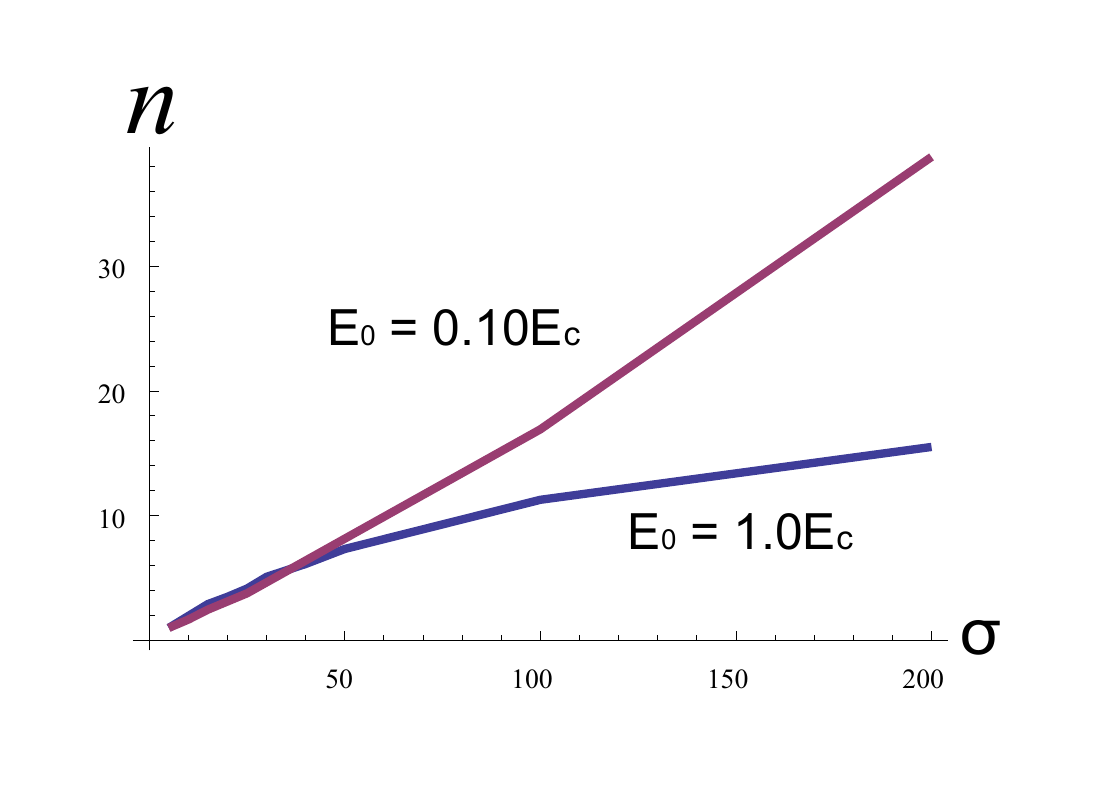} \hfill
\includegraphics[width=0.48\textwidth]{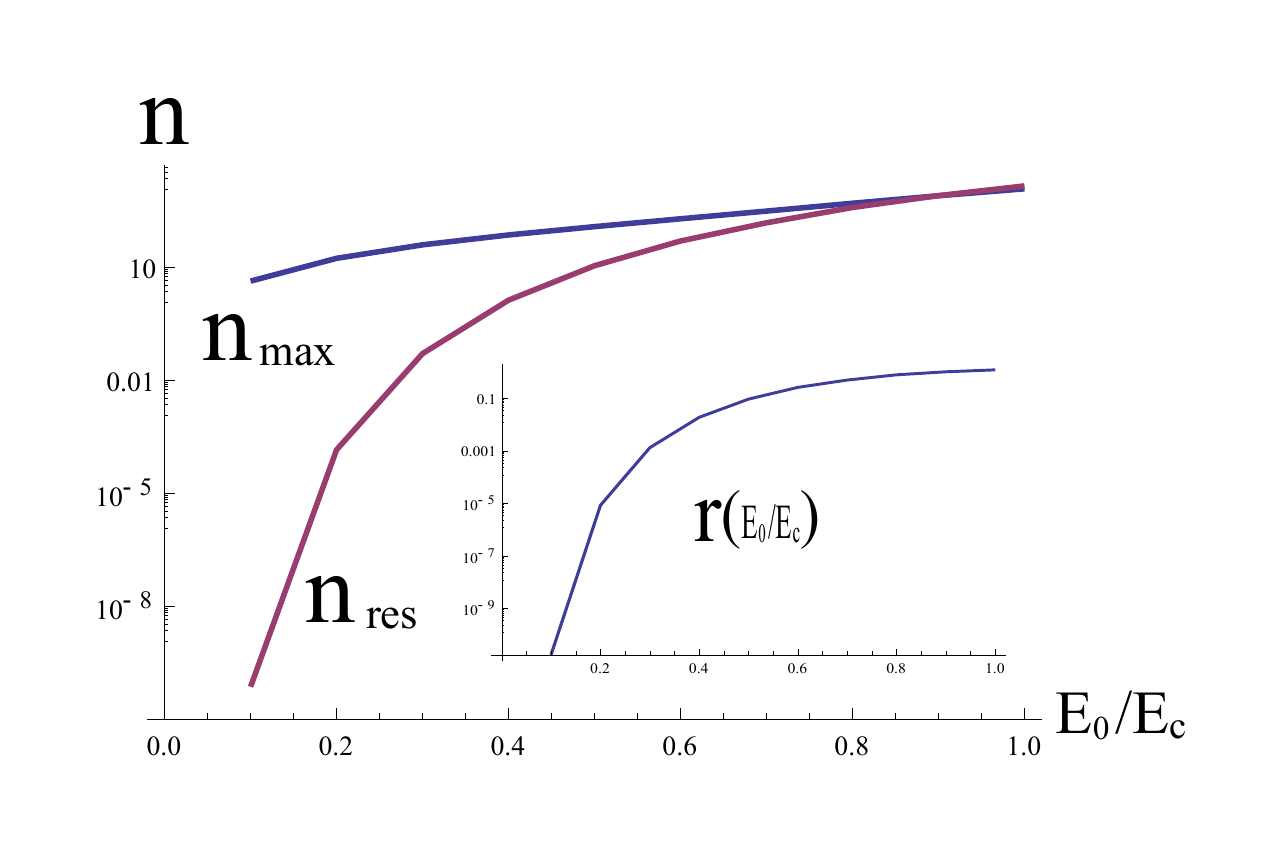}
\caption{Influence of the parameters the field pulse on REPP production.
{\bf Left panel:}  Influence of the laser pulse (\ref{field3}) duration $\sigma$ on the REPP density for the medium ($E_0/E_c=0.1$) and high ($E_0/E_c=1.0$) amplitude of the electric field. For both values of the field the unit value of density is determined as the density of the REPP produced by a pulse with a duration $\sigma = 5$.
{\bf Right panel:} Comparison of the maximum density of QEPP and density of the REPP in the range of values $0.1E_c \le E_0 \le 1.0E_c$ for the laser pulse (\ref{field3}) with a wavelength $0.02426~$nm 
($\omega = 0.1$) and $\sigma = 5$. The insertion shows the transformation coefficient $r(E_0/E_c) =n_{res}/ n_{max}$ for the shown values of the densities.
\label{fig:7}}
\end{figure*}

The left panel of Fig.~\ref{fig:7} demonstrates the dependence of the EPP pair density (\ref{dens}) on the pulse duration at fixed frequency $\omega$ in the field model (\ref{field3}). 
It exhibits a nonlinear accumulation effect for which the slope is approximately constant for weak fields 
whereas for strong fields we observe a saturation effect. 
Finally, the right panel of Fig.~\ref{fig:7} shows the dependence of the EPP pair density in the out-state in comparison to the maximal value attained within the entire period of the EPP evolution (see, e.g., Fig.~\ref{fig:1}, right).

\subsection{Strong nonequilibrium}

The entire process of vacuum EPP creation is a strong nonequilibrium one, including the final out-state.
In the first place, this conclusion follows from the exactly solvable models.
The distribution functions  of the out-state turn out to be the same for both, the constant field model $E(t) = E_0$ \cite{Cooper_1993} and the Eckart-Sauter model (\ref{field2}) for $T \to \infty$ \cite{Gavrilov_1996, Grib_1994}
\begin{equation}
\label{f_degen}
f_{\rm out}(\mathbf{p}) = \exp \left[-\frac{E_c}{E_0} \left(\frac{\varepsilon_{\bot}}{m}\right)^2 \right].
\end{equation}
This function is degenerate w.r.t. $p^3 = p_\parallel$ and therefore non-normalizable.
This leads to the necessity to extend the definition of macroscopic observables of the type (\ref{dens}).
As a rule, the substitution
\begin{equation}
\label{substitution}
   \int dp_{\parallel} \to eTE_0
\end{equation}
is introduced which results in the well known Schwinger formula \cite{Schwinger_1951} for the EPP production rate. 
The constant field model has been analyzed in detail in the recent work \cite{Tanji_2009}.

The strongly anisotropic nonequilibrium distribution (\ref{f_degen}) exists only in the presence of the external field and is defined by its symmetry.
Detailed consideration of the nonequilibrium feature of this distribution can found in the work \cite{Spokoiny_1982}.

\begin{figure*}
\includegraphics[width=0.48\textwidth]{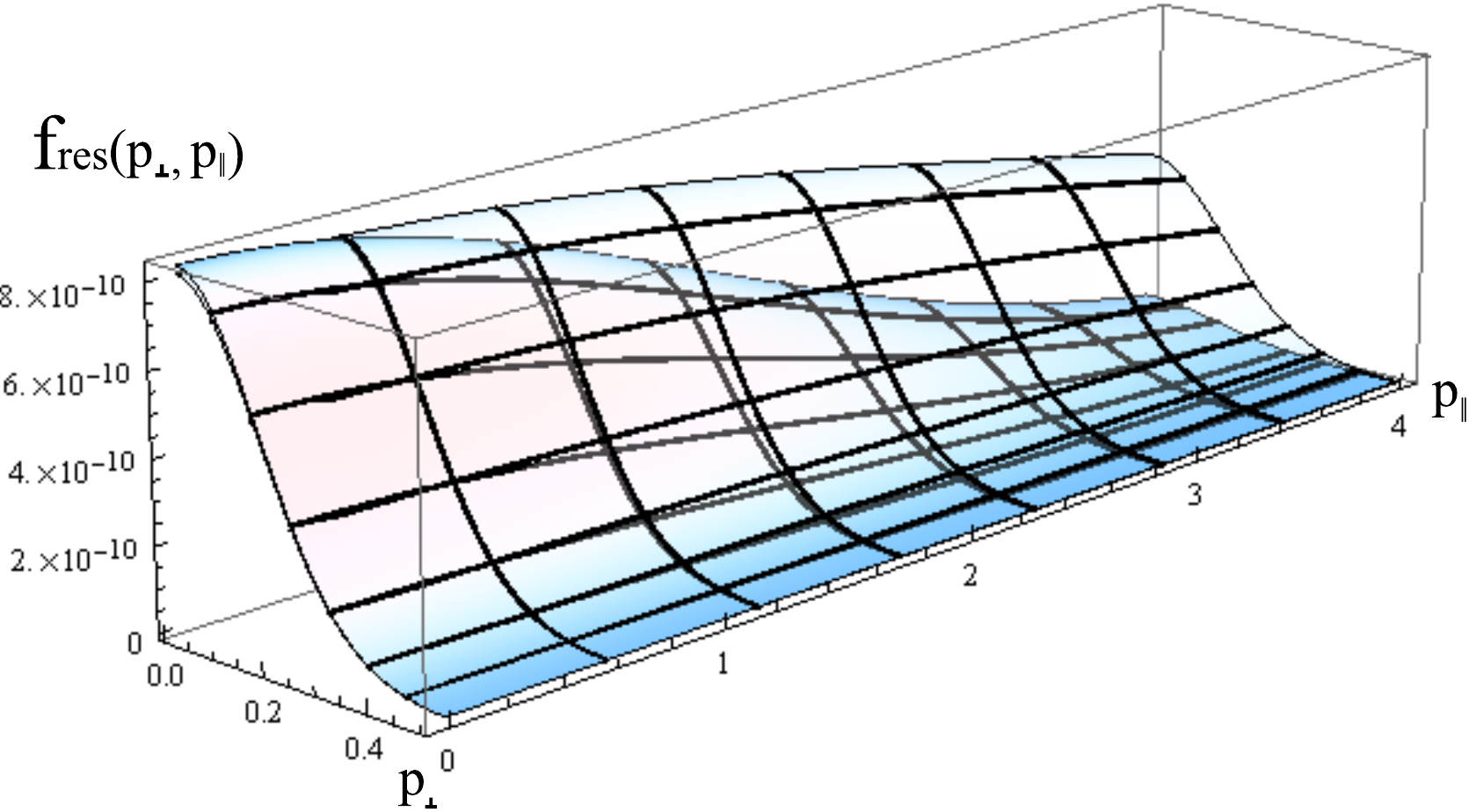} \hfill
\includegraphics[width=0.44\textwidth]{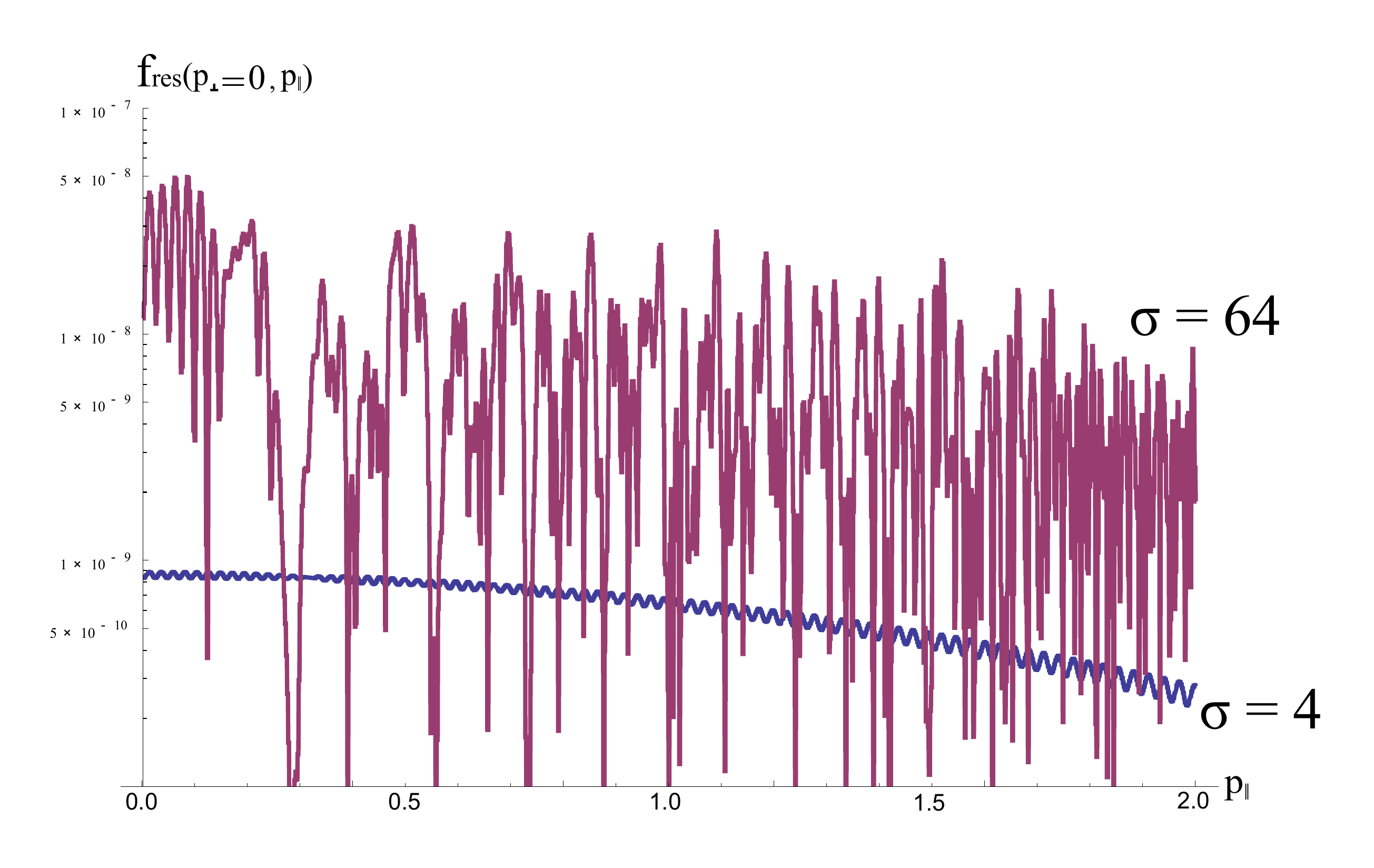}
\caption{
{\bf Left panel:} REPP distribution function $f(p_\bot,p_\parallel)$ for the Eckart-Sauter pulse (\ref{field2})  with $T = 82.4$ (bottom surface) and with $T = 164.8$ (top surface) at $E_0/E_c=0.15$ in momentum space for $0.0 \leq p_\bot \leq 0.5, 0.0 \leq   p_\parallel \leq 4.0$. {\bf Right panel:}  REPP distribution functions $f(p_\bot=0,p_\parallel)$ for a short pulse $\sigma = 4$ and a long pulse $\sigma = 64$ with the same amplitude of the field $E_0 = 0.15 E_c$. 
The cyclic frequency $\omega$ of the oscillating field satisfies the condition $1/\omega = 82.4$ that corresponds to the wavelength $0.1 nm$.\label{fig:6}}
\end{figure*}

The asymptotic distribution (\ref{f_degen}) in the constant field model is a smooth function of the transversal energy $\varepsilon_\bot(p_\bot)$.
In more realistic field models the structure of the distribution function becomes very complicated. 
As an example, see the right panel of Fig.~\ref{fig:6}.

\subsection{Non-monotonic entropy growth}

The transition from the in-state to the out-state is accompanied by an non-monotonic entropy growth.
This phenomenon was marked and discussed long ego (e.q., \cite{Rau_1996, Cooper_1993, Habib_1996}).
For example, the function (\ref{f_degen}) leads to the following entropy production rate 
\begin{equation}\label{S_v}
\frac{S_{\rm out}}{T} = \frac{m^4}{8\pi^2} \frac{E_0}{E_c} \left( 1+ \frac{E_0}{\pi E_c }\right) 
\exp \left(-\pi \frac{E_c}{E_0} \right),
\end{equation}
where the pulse duration is defined by the relation (\ref{substitution}). 
In Eq.~(\ref{S_v}) the definition of the information entropy 
with the density $s({\bf p},t) = - \ln f({\bf p},t)$ was used.
The most complete investigation was implemented  in the work \cite{Smolyansky_2012} on the basis of 
the KE (\ref{ke}).
Let us notice that the KE  (\ref{ke}) or the system of ODE (\ref{ode}) is invariant with respect to time inversion so that the entropy growth observed here, apparently, results from transforming the primordial vacuum fluctuations under the action of a strong external field to the statistical ensemble of the EPP with well defined entropy.

\section{Summary \label{sect:3}}

In this work, we have considered the field induced phase transition from primordial vacuum fluctuations to the final massive quantum field system of particle-antiparticle pairs under the action of a strong external field. 
This phenomenon possesses the following characteristic features
\begin{enumerate}[nolistsep]
\item[(i)] presence of three stages of evolution: quasiparticle, transient and final;
\item[(ii)] presence of fast oscillations in the transient stage;
\item[(iii)] strong nonequilibrium character, including the out-state;
\item[(iv)] non-monotonic entropy growth.
\end{enumerate}

Apparently, these features are rather universal and are characteristic on the qualitative level for physical systems of different nature.

On the formal level this universality appears because the corresponding KE's belong to the united class of integro-differential equations of non-Markovian type with fastly oscillating kernel.
Examples of this kind are, e.g., KE's for description of the vacuum creation of scalar bosons and of fermions in the FRW space-time \cite{Grib_1994}, the noncontradictory KE for massive vector bosons in the same metric \cite{Dmitriev_2017} and the nonperturbative KE for description of the carrier exitations in graphene \cite{Panferov_2017}.

In the present work we have restricted ourselves to the consideration of the domain of the tunneling mechanism of particle creation, $\gamma \ll 1$. 
We plan to consider the few-photon domain of particle creation ($\gamma \gg 1$) in a separate work.

\section{Acknowledments
}
D.B. and S.A.S. would like to thank A. Fedotov and S. Nedelko for their interest in this work and for useful discussions during a visit at JINR Dubna.
The work of D.B., L.J., and S.A.S. was supported in part by a grant from NCN under contract number UMO-2014/15/B/ST2/03752. L.J. is grateful for support of his visits to JINR Dubna from the Bogoliubov-Infeld program for collaboration between JINR Dubna and Polish research centres. 
S.A.S. acknowledges support for his participation as a lecturer in the Helmholtz International Summer School (HISS) on "Quantum Field Theory at the Limits: from Strong Fields to Heavy Quarks" in Dubna, July 18-30, 2016, where this work was completed.

\end{document}